\documentclass[a4paper,11pt]{article}

\usepackage{color}

\usepackage[disable]{todonotes}

\usepackage[T1]{fontenc}
\usepackage[utf8]{inputenc}
\usepackage[italian,english]{babel}
\usepackage{amsmath}
\usepackage{amssymb}
\usepackage{braket}

\title{Identical particles exchange symmetry\\ and the electric dipole moment in molecules.}

\author{Guglielmo M. Tino\\
 Dipartimento di Fisica e Astronomia and LENS Laboratory\\
Universit\`a di Firenze, INFN, CNR-INO\\ via  Sansone 1,  Sesto Fiorentino, Italy\\ 
email: guglielmo.tino@unifi.it}

\begin{document}
\maketitle

\begin{abstract}

%

Based on fundamental symmetries,  molecules cannot have a permanent electric dipole moment although it is commonly used in the literature to explain the different molecular spectra for heteronuclear and homonuclear molecules.
Electric-dipole rotational and vibrational spectra can indeed be observed  in heteronuclear molecules while they are missing in molecules with identical nuclei. I show that the missing spectral features can be explained as an effect of the exchange symmetry for identical particles.

\todo{verifica risultati testo antiplagio?}

\end{abstract}

%
%
%
%
%

%
%

%
%
\section{Introduction}
\label{sec:Intro}
%
%

Papers discussing the electric dipole moment of fundamental particles   state that atoms and molecules cannot have a permanent electric dipole moment unless fundamental symmetries, namely,  time reversal ($T$) and parity ($P$) are violated (see, for example, \cite {Bernreuther1991,Chupp2019}). 
Such violations  are searched for in ongoing experiments on the electron's electric dipole moment 
\cite{Commins2007,DeMille-LesHouches,Cairncross2017,Andreev2018}. 

On the other hand, most texts discussing molecular spectra  state that electric-dipole vibrational and rotational transitions can  be observed in heteronuclear molecules that would have such a permanent electric dipole moment (see, for example, \cite {Herzberg_radicals,Townes-Schawlow,Atkins-Friedman,Bransden-Joachain}) while they cannot be observed in homonuclear molecules because they cannot have a permanent electric dipole.

The scope of this note is to reconcile these two drastically opposing statements on fundamental symmetries and clarify that molecules cannot have a permanent electric dipole moment and nevertheless electric-dipole rotational and vibrational spectra can  be observed in heteronuclear molecules but they are missing in  molecules with identical nuclei as an effect of the exchange symmetry.  

It should be noted that I refer to  molecules in their nondegenerate ground state; permanent electric  dipole moments were indeed investigated for homonuclear molecules with asymmetric electronic excitation between the atoms (\cite{Li2011,Lu2022} and references therein).

A simple model for neutral diatomic molecules is considered here but the main conclusion is valid  also  for more complex molecules including finer effects \cite{Klemperer1993}.

%
%
\section{The electric dipole moment in molecules.}
\label{sec:EDM}
%
%

The permanent electric dipole ${\bf D}$ of a molecule  in a state $\ket{\psi}$ is:
\begin{equation}
{\bf D} =  \bra{\psi}{\bf D}\ket{\psi},
\label{DD}
\end{equation}
where the electric dipole operator ${\bf D}$  in the nonrelativistic limit is given by 
%
\begin{equation}
 {\bf D} = e \left(\sum_{n}Z_n{\bf R}_n - \sum_{m}{\bf r}_m \right),
\label{D}
\end{equation}
with $e$  the elementary charge,  the first sum running over the  positions ${\bf R}_n$ of the nuclei with  charges $Z_n e$, and the second sum  running over the positions ${\bf r}_m$ of the electrons 
\footnote{The electric dipole moment operator is independent of the choice of the origin for neutral systems.}.

The typical argument on a possible permanent electric dipole moment in molecules and the resulting rotational and vibrational spectra goes as follows \cite{Bransden-Joachain}: 
%
%
In the case of atoms, the permanent electric dipole moment is always zero for non-degenerate states because they are eigenstates of the parity operator $P$ and {\bf D} is odd under this transformation. The permanent electric dipole moment would be instead different from zero for heteronuclear molecules, such as $\rm HCl$, because an excess charge would be associated to one of the nuclei compared to the other. In symmetrical homonuclear diatomic molecules, such as $\rm H_2, O_2, N_2$, the permanent electric dipole moment is instead zero. 
As a consequence, rotational and vibrational spectra would only be observed for heteronuclear molecules because the transition amplitude for absorption or emission of radiation by a molecule  is proportional (in the electric-dipole approximation) to the matrix element of the electric dipole operator. 

This is consistent indeed with what is observed experimentally and no such spectra can be observed for homonuclear  molecules but the explanation reported above is, in my opinion, misleading because molecules cannot have a permanent electric dipole moment. 

Let's clarify how a permanent electric dipole moment in a molecule would manifest itself and the relation with molecular spectra \cite{Bernreuther1991,DeMille-LesHouches}. 

In the presence of an external electric field $ \vec{\mathcal{E}}$, the interaction energy of a neutral particle with the field is
\begin{equation}
\Delta {\rm E} = \sum_{i}{\rm D}_i \mathcal{E}_i + \sum_{i,j}{\rm \alpha}_{ij}\mathcal{E}_i \mathcal{E}_j + ...
\label{DeltaE}
\end{equation}
The first term is linear in $\mathcal{E}$ and it is non null if a static electric dipole moment ${\bf D}$ is present. The second term, that depends on the polarizability tensor ${\rm \alpha}_{ij}$, is instead due to an induced electric dipole moment which is produced by the field and interacts with the field itself; in this case, the dependence on the applied electric field is quadratic. 

So,  the presence of a permanent electric dipole in a system manifests itself with a change in energy $\Delta {\rm E}$ proportional to the applied field $\mathcal{E}$. 


%

A permanent electric dipole moment for a particle at rest should  be proportional to the particle spin or total angular momentum that provide a reference direction; 
but 
${\bf D}$ is even under time-reversal $T$ and  odd under parity-reflection $P$ 
while 
spin is odd under $T$ and even under $P$. 
Therefore, 
an elementary or composite particle, such as an atom or a molecule, cannot have a permanent electric dipole moment unless  $T$ and  $P$ symmetries are violated \cite {Bernreuther1991,Chupp2019}. 

This is the rationale for the ongoing experimental search of an electric dipole moment in atoms and molecules (see \cite{Commins2007,DeMille-LesHouches,Cairncross2017,Andreev2018} and references therein). In these experiments, one searches for a linear dependence of the energy on an applied electric field in a paramagnetic atom or molecule with one or more unpaired electrons
looking for an electric dipole moment proportional to the spin of the system, so that when the spin is reversed so would be the $P,T$-violating electric dipole moment.
Since a linear  effect corresponds to a permanent atomic or molecular electric dipole moment, the  null results obtained so far in the experiments are interpreted as an upper bound to a possible electric dipole moment of the unpaired electron.



The reason for the misleading statement about a permanent electric dipole moment of heteronuclear molecules stems from the presence of quasi-degenerate states with opposite parities that can be mixed by an external electric field thus simulating the presence of a dipole moment. 

As is well known in the case of the hydrogen atom,  considering as degenerate the excited states with different $l$ values for the orbital angular momentum,  the presence of an external electric field mixes states with opposite parities given by $(-)^l$; this leads to a linear dependence of the energy change on the applied electric field and we can say that in this state the atom has a permanent electric dipole moment. 

This is strictly true, however, only if the opposite-parity states have the same energy. If the opposite-parity states are quasi-degenerate, for weak electric fields the energy dependence is given by the second-order term that is quadratic in the field and corresponds to an induced electric dipole moment; only when the effect of the electric field is large enough to make the energy separation between the unperturbed levels negligible, the linear behaviour is approached. So, the smaller the energy separation between nearby opposite-parity states, the larger the polarizability and the better the system can approximate the behaviour of a permanent electric dipole moment.

In the case of molecules, this effect can be produced by the electric field mixing nearby rotational states which have a parity given by $(-)^K$, where $K$ is the rotational quantum number. However, as in the case of the hydrogen atom, this is not an evidence of the presence of a permanent electric dipole moment nor an indication of $P$ and $T$ violation; if a sufficiently weak electric field were applied, the quadratic dependence  would show the induced origin of the observed dipole moment. 

This clarifies the origin and the real meaning of the expression "permanent electric dipole moment" for heteronuclear molecules.

\section{Identical particles exchange symmetry and the missing electric dipole in homonuclear molecules}


Why the argument above does not work for homonuclear molecules and they do not show even an apparent "permanent electric dipole moment" nor electric-dipole rotational and vibrational spectra? 
That is indeed an effect of the exchange symmetry for identical particles.

According to the symmetrization postulate of quantum mechanics and the spin-statistic connection, 
if the nuclei are identical and have an integer (half-integer) spin, 
the total molecular wave function $\psi_t$ must be symmetric (anti-symmetric) in the exchange of two nuclei. 

 The total  wave function $\psi_t$ can be written, in the Born-Oppenheimer approximation and neglecting the very small coupling of the nuclear spin with the rest of the molecule, in the form 
%
\begin{equation}
\psi_t = \psi_e\psi_v\psi_r\psi_n,
\label{psi}
\end{equation}
where $\psi_e$, $\psi_v$, and $\psi_r$ are the electronic, vibrational, and rotational functions, respectively, and $\psi_n$ is the nuclear spin function.

The electronic wave function $\psi_e$ can be either even or odd under the exchange of the nuclei. 
The vibrational wave function $\psi_v$ is  symmetric in the exchange of the nuclei because it depends only on the magnitude of the internuclear distance. 

Considering the rotational states, if the nuclei have no nuclear spin (I=0), alternate rotational states with even or odd rotational quantum numbers $K$ are forbidden, depending on the symmetry of $\psi_e$, because of their parity  $(-)^K$. 
As a result, the mixing by an electric field of nearby opposite-parity rotational levels is not possible in this case. 
This is the case, for example, of the $\rm ^{16}O_2$ molecule for which  high-sensitivity  spectroscopy experiments were performed searching for the forbidden states as an indication of possible violations of the symmetrization postulate of quantum mechanics and of the spin-statistic connection (see \cite{deAngelis1996,Tino2000,Hilborn-Tino2000} and references therein). 

If the nuclei have a nuclear spin, different rotational levels will be associated to different values of the total nuclear spin. In the case of $\rm ^1H_2$ ($\rm I=1/2$), for example,  alternate rotational states are associated to singlet and triplet nuclear spin states; the transitions between these states are extremely weak \cite{Pachucki2008} so they are virtually uncoupled.

Therefore, in the case of molecules with identical nuclei,  the mixing by an electric field of nearby opposite-parity rotational levels is virtually impossible because of the exchange symmetry; thus  there is no linear dependence of the energy shift on the electric field and no apparent "permanent electric dipole moment" in homonuclear molecules. 

\section{Conclusions} 
\label{sec:Concl}

In conclusion,  molecules, either homonuclear or heteronuclear, cannot have a permanent electric dipole moment because of $P$ and $T$ symmetries. The apparent behaviour  as if a permanent electric dipole moment was present in heteronuclear molecules is correctly explained in terms of the mixing of rotational states induced by an external electric  field. This mixing is not possible  in homonuclear molecules due to the exchange symmetry for the identical nuclei; therefore in this case there is not even an apparent electric dipole moment.

\vspace{1cm}
{\bf Acknowledgements}

The author thanks 
J.~Burgd\"orfer, D.~P.~DeMille, K.~Pachucki for the critical reading of a preliminary version of the manuscript and A.~Blech,   L.~Kerber,  M.~Leibscher,   A.~S.~Tino for useful discussions.
 \\



\end{document}